# Topological aspects of QCD.

A. Di Giacomo[*] [a]

[a]Dipartimento di Fisica Università di Pisa and INFN Sezione di Pisa

We review recent results from lattice on topological aspects of QCD: most of the results refer to monopoles and to instantons. We discuss in detail the evidence for condensation of monopoles in the vacuum and confinement of colour by dual superconductivity, and the major role of monopoles in dynamics (monopole dominance). As for instantons we review the $U(1)$ problem, a possible determination of the spin content of the proton, and new lattice data relevant to instanton liquid models.

## 1. Introduction

The role of configurations with non trivial topology in gauge theories has been extensively studied on the lattice. The main part of this talk will be a review of recent work on monopoles and on instantons.

In sect. 2.1 we will review recent evidence for the condensation of monopole charges in the vacuum, producing dual superconductivity and confinement of colour. Increasing evidence is also being collected of the major role that monopoles play in determining physical quantities like the string tension, the chiral condensate, the masses of hadrons. This property is known as monopole dominance and will be reviewed in sect. 2.2. Both sect. 2.1 and sect. 2.2 will cover results subsequent to the last review on the subject at LAT92[1].

As for instantons recent progress relevant to the $U(1)$ problem will be presented in sect. 3.1: this will mainly concern the program of improving the action or the operator for the topological charge or both, in order to reduce lattice artifacts. Some progress on the study on the lattice of the spin content of the proton appeared after LAT94[2] will be discussed in sect. 3.2.

Sect. 3.3 will cover advances on the understanding of the dynamical relevance of instantons after the review at LAT93[3].

Sect. 4 will rapidly mention some new seminal ideas related to topology appeared in the literature.

Concluding remarks will follow in sect. 5.

## 2. Monopoles

### 2.1. Condensation of monopoles and colour confinement

One of the most appealing mechanisms for confinement of colour is type II dual superconductivity (SC) of the vacuum[4–6]. Dual means interchange of the roles of electric and magnetic quantities with respect ordinary SC. The idea is that dual Meissner effect squeezes the chromoelectric field acting between a $q\,\bar{q}$ pair into an Abrikosov flux tube, producing a force proportional to the distance.

Ordinary SC is a $U(1)$ Higgs phenomenon[7]. The charged field $\Phi$, describing the annihilation of a Cooper pair acquires a non zero vacuum expectation value (v.e.v.), $\langle\Phi\rangle \neq 0$ thus breaking spontaneously the $U(1)$ symmetry related to charge conservation. Parametrizing the field $\Phi$ by its modulus $\psi$ and phase $\Phi = \psi e^{iq\theta}$ and defining the gauge invariant quantity $\tilde{A}_\mu = A_\mu - \partial_\mu \theta$, one has $F_{\mu\nu} = \partial_\mu \tilde{A}_\nu - \partial_\nu \tilde{A}_\mu$. The equation of motion for the photon is, neglecting fluctuations of the Higgs field,

$$\partial_\mu F_{\mu\nu} + m^2 \tilde{A}_\nu = 0 \qquad m^2 = q^2 \langle\Phi\rangle^2 \qquad (1)$$

For a static configuration, with no charges, $F_{0i} = 0$ and Eq(1) gives

$$\vec{\nabla} \wedge \vec{H} + m^2 \vec{\tilde{A}} = 0 \qquad (\vec{H} = \vec{\nabla} \wedge \vec{\tilde{A}}) \qquad (2)$$

---

[*]Partially supported by MURST and by EC Contract CHEX-CT92-0051

Taking the curl of Eq(1)

$$(\nabla^2 - m^2)\vec{H} = 0 \qquad (3)$$

Eq(2) implies the presence of a persistent current

$$\vec{j} = m^2 \vec{A} \qquad (4)$$

(London current) with zero electric field, i.e. zero resistivity. Eq(3) implies that $\vec{H}$ dies off exponentially in the material (Meissner effect). The key parameter is $m^2$ or $\langle \Phi \rangle \neq 0$ which signals S.S.B. of $U(1)$, and is called order parameter. Dual superconductivity will be signalled by a non vanishing v.e.v. of a magnetically charged field $\langle \Phi_M \rangle$. $\langle \Phi_M \rangle$ is called a disorder parameter for reasons which will be clear below.

In non abelian gauge theories the generic monopole configuration is identified, up to a gauge transformation by an $x$ independent matrix of the algebra, diagonal, with integer eigenvalues[8,9] (GNO classification). This means that any hermitian or unitary local operator after diagonalization by a gauge transformation identifies a monopole species, and conversely any monopole species identifies a set of commuting operators which are made diagonal with the corresponding GNO matrix by a gauge transformation. Monopoles can be stable only for groups with non trivial first homotopy group $\Pi_1(G)$[9,10]. Since $\Pi_1(SU(N)) = 1$ no stable monopoles exist in $SU(N)$ gauge theories. A stable monopole exists for $SO(3)$ for which $\Pi_1(SO(3)) = Z_2$ and is known as 't Hooft - Polyakov monopole[11,12]. The model is a Higgs model with gauge group $SO(3)$ with the scalar field belonging to the vector representation. In the broken phase a static monopole configuration with finite energy exists. A gauge invariant field strength can be defined[11]

$$F_{\mu\nu} = \hat{\vec{\Phi}} \cdot \vec{G}_{\mu\nu} + \frac{1}{g} \hat{\vec{\Phi}} \cdot \left( D_\mu \hat{\vec{\Phi}} \wedge D_\nu \hat{\vec{\Phi}} \right) \qquad (5)$$

where $\hat{\vec{\Phi}} = \vec{\Phi}/|\vec{\Phi}|$. If we define $B_\mu = \hat{\vec{\Phi}} \cdot \vec{A}_\mu$, $B_\mu$ is not gauge invariant since $\vec{A}_\mu$ is not covariant. The identity holds

$$F_{\mu\nu} = (\partial_\mu B_\nu - \partial_\nu B_\nu) + \frac{1}{g} \hat{\vec{\Phi}} \cdot \left( \partial_\mu \hat{\vec{\Phi}} \wedge \partial_\nu \hat{\vec{\Phi}} \right) \qquad (6)$$

Choosing a gauge such that $\hat{\vec{\Phi}} = (0, 0, 1)$ makes the last term in Eq(6) is zero and

$$F_{\mu\nu} = \partial_\mu B_\nu - \partial_\nu B_\nu \qquad (7)$$

This gauge is determined up to a $U(1)$ gauge rotation around the 3-rd axis, and is called an abelian projection. The abelian projection identifies the matrix which classifies the monopole according to GNO. The gauge invariant field strength of the monopole at large distances is a Dirac monopole: $\vec{E} = 0$, $\vec{H} = \vec{r}/(gr^3)$. Monopole charge belongs to a $U(1)$ which is not a subgroup of the original gauge group, is gauge invariant, and coincides with a subgroup only in a particular gauge, i.e. after abelian projection.

Dual superconductivity of the vacuum will occur if some monopoles species (monopoles defined by some abelian projection) condense in the vacuum.

Popular choices for the abelian projection are

(1) The projection which diagonalizes the Polyakov line $P(n) = \prod_{i=1}^{L_T} U_0(n + i\hat{n}_0)$.

(2) The projection which diagonalizes a component of the field strength, say $F_{12}(n)$: on the lattice this means the open plaquette $\Pi_{12}(n)$ (no trace).

(3) The maximal abelian projection identified by the gauge transform $\Omega$ defined by a maximization[13]

$$\text{Max}_\Omega \sum_{n,\mu} \text{Tr} \left\{ \sigma_3 \Omega(n) U_\mu(n) \right.$$
$$\left. \Omega^\dagger(n+\mu) \sigma_3 \Omega(n+\mu) U^\dagger(n) \Omega^\dagger(n) \right\}$$

In this case the operator which is diagonal is not defined explicitly.

't Hooft guess[14] is that may be all species of monopoles do condense and are relevant for confinement. The basic question is: what monopoles do condense, if any.

**2.2. Defining monopole currents.**

In $U(1)$ gauge theory the plaquette in the $\mu$, $\nu$ plane is $\Pi_{\mu\nu} = e^{i\theta_{\mu\nu}}$ with $\theta_{\mu\nu} = \sum_{links} \theta_i$. Since

$-\pi \leq \theta_i \leq \pi$, we have $-4\pi \leq \theta_{\mu\nu} \leq 4\pi$. Define[15] $\overline{\theta}_{\mu\nu}$ as $\theta_{\mu\nu}$ mod$2\pi$,

$$\overline{\theta}_{\mu\nu} = \theta_{\mu\nu} + 2n\pi n_{\mu\nu} \tag{8}$$

with $-\pi \leq \overline{\theta}_{\mu\nu} < \pi$. $n_{\mu\nu}$ counts the number of Dirac strings across the plaquette and is bounded by $-2 \leq n_{\mu\nu} \leq 2$. For a 3d cube with edges in the directions $\mu,\nu,\rho$ an integer $n_{\mu\nu\rho}$ can be defined, counting the net number of Dirac strings across external surface and $j_\alpha$ as $j_\alpha \varepsilon^{\alpha\mu\nu\rho} = n^{\mu\nu\rho}$. In particular $j_0$ is the space density of monopoles. The kinematic bound on $j_0$ is $-12 \leq j^0 \leq 12$. The definition of $j$ is sensible only if the density of monopoles is $\ll 12$, otherwise monopole density will depend critically on the size of the cube used to define it. In non abelian theories after abelian projection one can uniquely define an abelian link $u_\mu = e^{i\theta_\mu \sigma_3}$ and construct monopole currents with it. Polyakov and Field Strength abelian projections have a density of monopoles which is large, and hence affected by lattice artifacts. The density of maximal abelian projection is instead $\ll 1$[16] which makes this projection particularly clean when looking at the density of monopoles.

### 2.3. Detecting dual superconductivity.

The density of monopoles is not an order parameter for dual superconductivity. To detect dual superconductivity a disorder parameter $\langle \Phi_M \rangle \neq 0$ must be detected which is the v.e.v. of an operator carrying magnetic charge. (The density of monopoles is a neutral operator, commuting with monopole charge). Two approaches have been used for this

(i) detecting the London current.

(ii) direct measurement of $\langle \Phi_M \rangle$.

### (i) Detecting the London current.
The dual of Eq(2) is

$$\vec{\nabla} \wedge \vec{E} + m^2 \vec{A} = 0 \quad \text{or} \quad m^2 \vec{E} = \vec{\nabla} \wedge \vec{j} \tag{9}$$

$$m^2 \propto \langle \Phi_M \rangle$$

If the superconductor is type II an Abrikosov line has a thin central part $\sim 1/m_\Phi$ where SC is destroyed and a larger region of radius of order of the penetration length $1/m$ in which the system is superconducting and Eq(9) holds. The first attempt[17] to determine $m$ by analysing such region for $U(1)$ and $SU(2)$ showed that such a region does not exists. The system is at the border between type I and type II ($1/m_\Phi \simeq 1/m$) so that one has to look at the explicit solution for the flux tube[18] and try a fit of the data. Similar results were obtained for $SU(2)$ and $SU(3)$ in ref.[19], and more recently in ref.[20], in the maximal abelian gauge. The result is compatible with superconductivity. Since $\vec{\nabla} \wedge \vec{j}$ is measured from the monopole current the maximal abelian projection works better in view of eliminating lattice artifacts. However the derivatives of $\vec{j}$ are computed by differences on transverse sizes which are typically of 2 - 3 lattice spacing and systematic errors could affect the best fit. A contribution to this conference by the Bari group[21] gives a precise determination of the $x_T$ dependence of the parallel electric field in a flux tube, which fits the expected behaviour $E_\parallel(x_T) = m^2 K_0(mx_T)\Phi_0/2\pi$.

### (ii) Direct measurement of the disorder parameter $\langle \Phi_M \rangle$[22,23]

A mathematical proof of monopole condensation exists for $U(1)$ gauge theory with Villain action[24] and for $N = 2$ supersymmetric Q.C.D.[25]. The basic concept of these constructions is that of duality transformation, which was first introduced in the 2d Ising model in ref.[26]. The model can be described either in terms of $\sigma = \pm 1$, or in terms of kinks, $\sigma^* = \pm 1$ on the dual lattice. The duality relation is an equality between partition functions $K[(\sigma, \beta] = K[\sigma^*, \beta^*]$ $\beta^* \sim 1/\beta$. The strong coupling regime of the $\sigma$ description corresponds to the weak coupling for $\sigma^*$, and viceversa. The duality relation holds in the limit $V \to \infty$. In that limit $\langle \sigma^* \rangle \neq 0$ when $\langle \sigma \rangle = 0$, and $\langle \sigma^* \rangle = 0$ when $\langle \sigma \rangle \neq 0$, therefore $\langle \sigma^* \rangle$ is called a disorder parameter. The disorder parameter $\langle \sigma^* \rangle$ is $\neq 0$ in the limit $V \to \infty$, in the phase $\langle \sigma \rangle = 0$.

We know, however, that monopoles do condense in the $U(1)$ theory with Wilson action, even if we are not able to make use of the duality relation. In non abelian gauge theories we have to detect monopole condensation without any knowledge of the effective action for the abelian projected $U(1)$ degrees of freedom. A progress has been made towards direct detection of dual SC by the construction of an operator with non trivial monopole charge, whose v.e.v. can be used to monitor the S.B. of the magnetic $U(1)$ symmetry. The construction applies to $U(1)$, and can be used for Q.C.D., for the $U(1)$ identified by any abelian projection.[22,23]

A creation operator for monopoles can be defined as a shift of the field configuration $\vec{A}(\vec{x})$ by the vector potential produced by a monopole of magnetic charge $m$ located in $\vec{y}$, $\vec{b}(\vec{x} - \vec{y})$

$$\mu = \exp\left[i \int d^3x\, \vec{\Pi}(\vec{x},t) \frac{1}{e} \vec{b}(\vec{x} - \vec{y})\right] \quad (10)$$

where $\vec{\Pi} = \vec{E}(\vec{x}, t)$ is the conjugate momentum to $\vec{A}$, and $\vec{b}$ can be chosen as

$$\vec{b}(\vec{x} - \vec{y}) = \frac{m}{2} \frac{\vec{n} \wedge \vec{r}}{r(r - \vec{r} \cdot \vec{n})} \qquad \vec{r} = \vec{x} - \vec{y} \quad (11)$$

with the prescription that the Dirac string, which is put in the direction $\vec{n}$, must be removed. $\mu$ has monopole charge $m$. Indeed if $Q_M = \int d^3x\, \vec{\nabla}\vec{H}$ by use of canonical commutations relations une easily obtains:

$$[Q_M, \mu(t)] = m\mu(t) \quad (12)$$

A careful definition of normal ordering[22] brings to define as a disorder parameter

$$\langle \bar{\mu} \rangle = \frac{\langle \mu \rangle}{\langle \gamma \rangle} \quad (13)$$

with

$$\gamma = \exp\left[i \int d^3x\, \vec{\Pi}(\vec{x},t) \frac{1}{e} \vec{g}(\vec{x})\right] \quad (14)$$

$\gamma$ describes a shift of the field by $\vec{g}(\vec{x})$. If $\vec{\nabla} \wedge \vec{g}(\vec{x}) = 0$ this shift does not modify the magnetic field, nor the electric field, since $\partial \vec{g}/\partial t = 0$, and $\gamma$ is a pure gauge transformation. $\vec{g}(x)$ is subjected to the condition

$$\int d^3x\, \vec{g}^2(x) = \int d^3x\, \vec{b}^2(x)$$

The correlation function of a monopole antimonopole pair $\langle \mu \bar{\mu} \rangle$ or any other correlator between any number of monopoles and antimonopoles can be defined in a similar way. It proves convenient to compute $\langle \bar{\mu} \rangle$ in terms of $\rho$

$$\langle \bar{\mu} \rangle = \exp \int_0^\beta \rho(x)dx$$

with $\rho = d\ln\langle \bar{\mu} \rangle / d\beta$. Explicitely one finds[22]

$$\rho = \langle S + S_g \rangle_{S+S_g} - \langle S + S_b \rangle_{S+S_b} \quad (15)$$

where $S$ is the action,

$$S_b(t) = \int d^3x\, \vec{\Pi}(\vec{x},t)\left[\sum_i \vec{b}(\vec{x}, \vec{y}_i)\right] \quad (16)$$

$$S_g(t) = \int d^3x\, \vec{\Pi}(\vec{x},t)\vec{g}(\vec{x}) \quad (17)$$

with

$$\int d^3x\, \vec{g}^2(\vec{x}) = \int d^3x\, \left[\sum_i \vec{b}(\vec{x}, \vec{y}_i)\right]^2 \quad (18)$$

The sum in Eq.(16-18) is extended to all monopoles and antimonopoles to be correlated. The disorder parameter $\langle \bar{\mu} \rangle$ can be defined, by the cluster property as

$$\langle \bar{\mu}(x)\bar{\mu}(y) \rangle \simeq_{|x-y|\to\infty} \langle \bar{\mu} \rangle^2 \quad (19)$$

It can be proved explicitly that $\langle \bar{\mu} \rangle \to 0$ as $V \to \infty$ in the free photon phase. For low $\beta$'s $\rho$ becomes independent of $V$ as $V \to \infty$, and shows a sharp negative peak at the deconfining phase transition which corresponds to a sharp drop of $\langle \bar{\mu} \rangle$. Fig.1 shows a characteristic shape of $\rho$ for Wilson $U(1)$.[22]

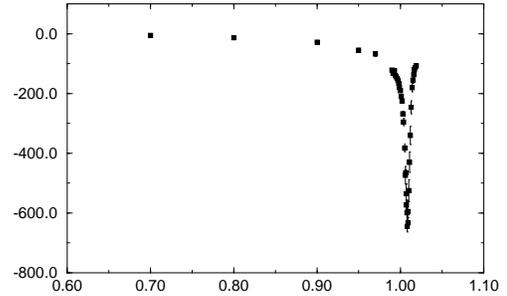

**Fig.1** $\rho$ vs $\beta$ on a lattice $8^4$.

A finite size scaling analysis[27] shows that the volume dependence in the region of the peak obeys the scaling law

$$\rho L^{-1/\nu} = f(L^{1/\nu}(\beta_c - \beta)) \qquad \mu \simeq_{\beta \to \beta_c} (\beta_c - \beta)^\delta$$

The scaling is shown in fig.2 and is well described by

$$\nu = 0.253(5) \quad \delta = 2.02(3) \quad \beta_c = 1.01099(5)$$

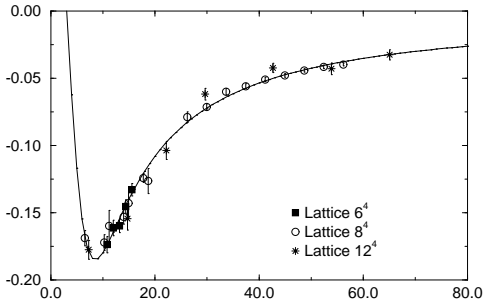

**Fig.2** $\rho/L^{1/\nu}$ vs $(\beta_c - \beta)L^{1/\nu}$

The value of $\nu$ indicates that the transition is first order and $\beta_c$ is consistent with other determinations[28,29].

For $SU(2)$ after abelian projection on the Polyakov line the behaviour of $\rho$ has a spectacular negative peak at the deconfining temperature[23]. A finite size analysis gives $\nu \simeq 0.65$ consistent with 3d Ising model, $\delta = 1.3 \pm .1$, $\beta_c(N_T = 6) - \beta_c(N_T = 4) = 0.048 \pm 0.002$[27].

For $SU(3)$ after abelian projection on the Polyakov line two monopoles charges can be defined[14]: the corresponding $\rho$'s have the same behaviour (Fig. 3) with a strong negative peak at the deconfining temperature[30].

Further study is needed to measure correlations, monopole mass, monopole effective potential but a clear demonstration emerges that $\langle \Phi_M \rangle \neq 0$, or that the vacuum is a dual superconductor. We stress again that the only way to detect superconductivity is to measure the v.e.v. of a charged operator: the density of monopoles cannot be a disorder parameter.

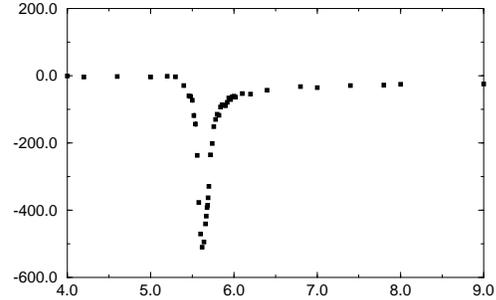

**Fig.3** $SU(3)$: $\rho$ vs $\beta$ on a lattice $12^3 \times 4$

### 2.4. Dynamical relevance of monopoles.

The idea of abelian dominance was already discussed in detail in ref.[1]. The statement is that abelian Wilson loops, defined after abelian projection, are a good approximation within 10% to the full Wilson loops.

This property has been extensively checked in the maximal abelian projection: more recently[31] it has been shown that it also works in the Polyakov loop abelian projections. Further work has shown that also the chiral condensate $\langle \bar{\psi}\psi \rangle$ is well approximated by abelian dominance in the maximal abelian projection[32,33], but not in the field strength projection[32].

Recently the concept of monopole dominance has been successfully introduced and explored. The idea is to separate in the abelian field the monopole contribution as defined by Eq(8) from the residual part. Calling $\theta_i$ the abelian phase of the $i$-th link, one solves $\theta_i$ in terms of $\bar{\theta}_{ij}$, $m_{ij}$. Symbolically in momentum space

$$\theta_i(k) \sim \frac{1}{k_j} \left( \bar{\theta}_{ij}(k) + \bar{m}_{ij}(k) \right) \qquad (20)$$

The relation Eq(20) is a convolution in space. Then one computes separately the contribution of monopoles ($m_{ij}$) and the Coulomb part ($\bar{\theta}_{ij}$), on physical quantities which are already known to be abelian dominated, and compares with the full loops. The construction goes back to ref.[34]. The overall result is that the Coulomb contribution is small and the monopole abelian part accounts for most of the physics. This is not unexpected, once abelian dominance is verified, since $m_{ij}$ is typically much larger than $\bar{\theta}_{ij}$ by construction: anyhow the results are impressive.

Monopoles do dominate

1) $\sigma$ in the max abelian[35,36] and Polyakov loop projection[31]

2) Fermion propagators and hadron masses[37,38].

3) The topological susceptibility[37].

It is apparent from Eq(20) that only large loops of monopoles currents (small $k$) do contribute to large Wilson loops: that is why monopole dominance of $\sigma$ extracted from large Wilson loops in the Polyakov loop abelian projection works in spite of the lattice artifacts which affect monopole density at short distances.

Much work has been done assuming dual SC and looking at observable consequences by use of effective lagrangians which I have no time to review. See e.g. ref.[39,40]

## 3. Instantons

### 3.1. The $U(1)$ problem

The existence of the anomaly of the $U(1)$ axial current in Q.C.D. provides a solution of the $U(1)$ problem.

$$\partial^\mu J_\mu^5 = 2N_f Q \qquad Q = \frac{g^2}{64\pi^2}\vec{G}_{\mu\nu}\vec{G}_{\mu\nu}^* \qquad (21)$$

The lattice community has widely contributed to clarify this problem. A measurement of the quenched topological susceptibility $\chi_a$ to verify the Witten Veneziano formula

$$\frac{2N_f}{f_\pi^2}\chi_Q = m_\eta^2 + m_{\eta'}^2 - 2m_K^2 \qquad (22)$$

has given for $\chi$ a value compatible with the expected value[41]

$$\chi \sim (180\,\text{MeV})^4 \qquad (23)$$

This result in fact tests at the same time Q.C.D. and the $1/N_c$ expansion.

An important result in the same direction is the measurement of the ratio

$$\frac{\langle \eta'(t)\eta'(0)\rangle_{2\,loops}}{\langle \eta'(t)\eta'(0)\rangle_{1\,loops}}$$

as a function of the topological charge of the configuration, $Q$. This ratio is an increasing function of $|Q|$ and the effective $\eta'$ mass is boosted by the presence of instantons[42].

The understanding of the measurements of the topological charge and susceptibility have definitely improved since the early days. In the quenched approximation lattice regularized version of the topological charge density $Q_L$ can only mix to the continuum $Q$ by multiplicative renormalization $Q_L = ZQ$. For the susceptibility there is a mixing to continuum susceptibility, to the action and to identity operator[43]

$$\chi_L = Z(\beta)\chi a^4 + M(\beta)G_2 a^4 + P(\beta) \qquad (24)$$

$Z$ and $M$ were missing in the so called naive approach. By cooling of quantum fluctuations $P(\beta) \to 0$, $M(\beta) \to 0$ and $Z(\beta) \to 1$ and $\chi_L \sim \chi a^4$[44]. A non perturbative determination of $Z$, $M$ and $P(\beta)$ can be performed[45]. By heating an instanton configuration $Z$ can be determined, and by heating the vacuum and measuring $\chi_L$, $P(\beta)$ and $M(\beta)$ can be determined[46].

Improved versions of $Q_L$ can be found[47] for which $Z \sim 1$ and $P(\beta)$ and $M(\beta)G_2 a^4$ are negligible with respect to $\chi a^4$ in a sizable scaling window.

A systematic improvement of the action, towards a classical perfect action is being pursued by the Bern group[48]: the aim is to extend the scaling window. The process implies a smearing of the action and of the operator on a few neighbouring sites: the goal is to reach a reasonable compromise between improvement and non locality. As for topology their aim is to put on safe grounds the so called geometrical definition of the topological charge by eliminating lattice artifacts (dislocations)[49]. This program is going on also for 2d asymptotically free spin models. Improving the operator and not the action[47] is worse from the point of view of scaling but can be easier in practice.

### 3.2. The spin content of the proton.

The matrix element of the $U(1)$ axial current on the nucleon is

$$\langle \vec{p}'s'|J_\mu^5|\vec{p}s\rangle = \bar{u}_{s'}(\vec{p}')\left[\gamma^5\gamma^\mu G_1(q^2) + G_2(q^2)\gamma^5 q_\mu\right]u(\vec{p})$$

$G_1(0)$ is related to the spin content of the proton. Naively one could use the anomaly equation Eq(21) and compute $G_1(0)$ as[2]

$$\lim_{q^2 \to 0} N_f \frac{\langle \vec{p}'s'|Q|\vec{p}s \rangle}{M_p \bar{u}_{s'}(\vec{p}')\gamma^5 u(\vec{p})}$$

In fact things are more complicated[50,51] since $Q$, $\partial^\mu J_\mu^5$ and $\bar{\psi}\gamma^5\psi$ mix by renormalization and $J_\mu^5$ has a small anomalous dimension. A correct treatment of the renormalization pattern can be done if $Q$ is defined by field theoretical formulae[43]. The computation makes sense only in full Q.C.D. (not quenched[2]) and appears feasible in particular by use of improved forms of $Q_L$.

### 3.3. Dynamical relevance of instantons.

After the report on this subject at LAT93, a nice result relevant to understanding of instanton liquid models is the study of the distribution in size of $SU(2)$ instantons and of the distribution in distance between $I$-$I$ and $I$-$\bar{I}$. The distribution shows, as expected, that instantons tend to concentrate at large sizes of the order of the correlation length, and that instantons of the same sign can coexist at shorter distances.

Similar results are beeing obtained with improved actions, which avoid deformation of instantons in the process of cooling[53].

### 4. New ideas.

I will finally rapidly quote a few papers with new ideas which could prove interesting.

Dyons could play a role in Q.C.D. dynamics[54].

The idea that the $\theta$ angle of strong interactions could be zero in the continuum limit for dynamical reasons, (phase structure of the theory) is attractive and has been tested on $CP_3$ model in 2d and $SU(2)$[55]. Maybe a safer determination of the topological susceptibility by a method different from the geometrical one would be welcome as a cross-check.

### 5. Outlook.

- There is increasing evidence that Q.C.D. vacuum is a dual superconductor. A more extensive study of monopole condensation on different abelian projections is necessary.

- Monopole dominance works impressively well.

- The $U(1)$ problem is well understood.

- The lattice determination of the spin content of the proton is at hand.